\documentstyle[twoside,fleqn,espcrc2]{article}

\newcommand{\AmS}{{\protect\the\textfont2
  A\kern-.1667em\lower.5ex\hbox{M}\kern-.125emS}}

\title{High energy neutrino astrophysics \hspace{60mm} {\small ADP-AT-98-8}}

\author{R.J. Protheroe\address{Department of Physics and Mathematical Physics\\
The University of Adelaide, Adelaide, Australia 5005}
}
       
\begin{document}

\begin{abstract}
I give a brief discussion of possible sources of high energy
neutrinos of astrophysical origin over the energy range from
$\sim 10^{12}$ eV to $\sim 10^{25}$ eV.  In particular I shall
review predictions of the diffuse neutrino intensity. Neutrinos
from interactions of galactic cosmic rays with interstellar
matter are guaranteed, and the intensity can be reliably
predicted to within a factor of 2.  Somewhat less certain are
intensities in the same energy range from cosmic rays escaping
from normal galaxies or active galactic nuclei (AGN) and
interacting with intracluster gas.  At higher energies, neutrinos
will definitely be produced by interactions of extragalactic
cosmic rays with the microwave background.  With the discovery
that gamma ray bursts (GRB) are extragalactic, and therefore
probably the most energetic phenomena in the Universe, it seems
likely that they will be copious sources of high energy
neutrinos.  Other sources, such as AGN and topological defects,
are more speculative.  However, searches for neutrinos from all
of these potential sources should be made because their detection
would have important implications for high energy astrophysics
and cosmology.
\end{abstract}

\maketitle


\section{INTRODUCTION}

The technique for constructing a large area (in excess of $10^4$
m$^2$) neutrino telescope has been known for more than two
decades \cite{BerZat77}.  The pioneering work of the DUMAND
Collaboration led to the development of techniques to instrument
a large volume of water in a deep ocean trench with strings of
photomultipliers to detect Cherenkov light from neutrino-induced
muons \cite{Learned91}.  Locations deep in the ocean shield the
detectors from cosmic ray muons.  The second generation of high
energy neutrino telescope such as AMANDA \cite{AMANDA} located
deep in the polar ice cap at the South Pole, and NT~200 in
operation in Lake Baikal, Siberia \cite{Baikal}, have
demonstrated the feasibility of constructing large area
experiments for high energy neutrino astronomy.  The next
generation telescopes, such as the planned extension of AMANDA,
ICECUBE \cite{ICECUBE}, and ANTARES \cite{ANTARES}, may have
effective areas of 0.1~km$^3$, or larger, and be sufficiently
sensitive to detect bursts of neutrinos from extragalactic
objects and to map out the spectrum of the diffuse high energy
neutrino background.  In this paper I focus on possible
astrophysical sources of neutrinos contributing to the diffuse
high energy neutrino background from $\sim 1$~TeV to the GUT
scale.

\section{COSMIC RAY INTERACTIONS WITH MATTER}

\begin{figure*}[htb]
\vspace{7.1cm}
\includegraphics{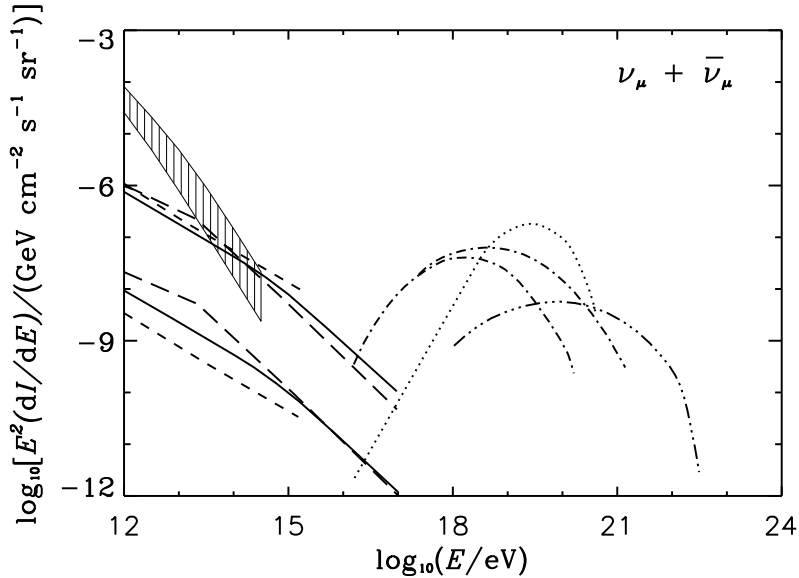}
\caption{Neutrinos from cosmic ray interactions with the
interstellar medium (upper curves for $\ell=0^\circ, \;
b=0^\circ$, lower curves for $b=90^\circ$): --- --- --- Domokos
et al. \protect\cite{Domokos}; - - - - Berezinsky et
al. \protect\cite{Berezinsky93}; -------- Ingelman and Thunman
\protect\cite{IngelmanThunman96}.  The band with vertical
hatching shows the range of atmosheric neutrino
background~\protect\cite{Lipari} as the zenith angle changes from
$90^\circ$ (highest) to $0^\circ$ (lowest).  Neutrinos from
cosmic ray interactions with the microwave background: $- \cdot -
\cdot - \cdot -$ Protheroe and Johnson
\protect\cite{ProtheroeJohnson95} for $E_{\rm max}=3\times
10^{20}$~eV and $3\times 10^{21}$~eV; $\cdots \cdots$ Hill and
Schramm \protect\cite{Hil85}; $- \cdot \cdot \cdot - \cdot \cdot
\cdot - \cdot \cdot \cdot -$ assuming the highest energy cosmic
rays are due to GRB according to Lee \protect\cite{Lee}.
\label{fig:cr3}}
\end{figure*}

There will definitely exist a diffuse galactic neutrino
background due to interactions of the galactic cosmic rays with
interstellar matter.  The spectrum of cosmic rays is reasonably
well known, as is the matter distribution in our galaxy.
Estimates of the neutrino intensity have been made by
Silberberg and Shapiro~\cite{Silberberg77},
Stecker~\cite{Stecker79}, Domokos et al.~\cite{Domokos},
Berezinsky et al.~\cite{Berezinsky93}, and Ingelman and Thunman
\cite{IngelmanThunman96}, and the more recent predictions are shown in
Fig.~\ref{fig:cr3}.  The differences of about a factor of 2
between the predictions are accountable in terms of the slightly
different models of the interstellar matter density, and cosmic
ray spectrum and composition used.  Also shown is the atmospheric
neutrino background as estimated by Lipari \cite{Lipari}.  In
addition, there will be a very uncertain background (not plotted)
due to charm production (see
refs. \cite{GaisserHalzenStanev95,Gaisser97} for a survey of
predictions).

\begin{figure}[htb]
\vspace{7.1cm}
\includegraphics{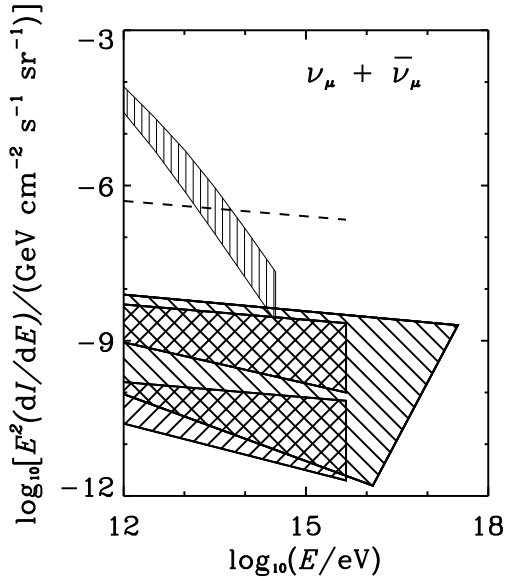}
\caption{Neutrinos from cosmic ray interactions in clusters of
galaxies (Berezinsky et
al. \protect\cite{BerezinskyBlasiPtuskin96}): lower hatched area
-- normal galaxies; upper hatched area -- AGN; dashed line --
upper bound from $\gamma$-ray data.  Large hatched region -- comic
ray interactions with intergalactic medium in clusters of
galaxies \protect\cite{ColafrancescoBlasi98}
\label{fig:cr2}}
\end{figure}

Somewhat less certain is the flux of neutrinos from clusters of
galaxies.  This is produced by $pp$ interactions of high energy
cosmic rays with intracluster gas.  Berezinsky et
al. \cite{BerezinskyBlasiPtuskin96} have made predictions of
this, and I show in Fig.~\ref{fig:cr2} their estimates of the
diffuse neutrino intensity due to interactions of cosmic rays
produced by normal galaxies and AGN together with an upper limit
based on assuming the observed $\gamma$-ray background results from
$\pi^0$ production.  Later estimates by Colafrancesco and
Blasi~\cite{ColafrancescoBlasi98} are also shown.

\section{COSMIC~RAY~INTERACTIONS WITH RADIATION}

Moving to higher energies, cosmic rays above $\sim 10^{20}$ eV
will interact with photons of the cosmic microwave background
radiation (CMBR) \cite{Gre66,Zat66}.  Again, we know that both
ingredients exist (the highest energy cosmic ray detected has an
energy of $3 \times 10^{20}$ eV \cite{Bir95}, and at least 6
cosmic rays have been detected above $10^{20}$ eV by the AGASA
array \cite{Takeda98}), and so pion photoproduction at these
energies will occur, resulting in a diffuse neutrino background
(Stecker~\cite{Stecker79}).  However, the intensity in this case
is model-dependent because it is not certain precisely what the
origin of the highest energy cosmic rays is, and whether in fact
they are extragalactic, although this seems very probable (see
\cite{ProtheroeERICE96CR} for a discussion of the highest energy
cosmic rays).  One of the most likely explanations of the highest
energy cosmic rays is acceleration in Fanaroff-Riley Class II
radio galaxies as suggested by Rachen and Biermann
\cite{RachenBiermann93}.  Protheroe and
Johnson~\cite{ProtheroeJohnson95} have repeated Rachen and
Biermann's calculation in order to calculate the flux of diffuse
neutrinos and $\gamma$-rays which would accompany the UHE cosmic
rays, and their result has been added to Fig.~\ref{fig:cr3}.  Any
model in which the cosmic rays above $10^{20}$ eV are of
extragalactic origin will predict a high energy diffuse neutrino
intensity probably within an order of magnitude of this at
$10^{19}$ eV.  For example, I show an earlier estimate by Hill
and Schramm \cite{Hil85}.  Also shown is an estimate by Lee
~\cite{Lee} of the diffuse neutrino intensity estimated in a
model in which the highest energy cosmic rays have their origin
in sources of gamma ray bursts.

\section{GAMMA RAY BURSTS}

Gamma ray bursts (GRB) are observed to have non-thermal spectra
with photon energies extending to MeV energies and above.  Recent
identification of GRB with galaxies at large redshifts (e.g.
GRB~971214 at $z=3.42$ \cite{Kulkarni98}) show that the energy
output in $\gamma$-rays alone from these objects can be as high as
$3 \times 10^{53}$~erg if the emission is isotropic, making these
the most energetic events in the Universe.  GRB~980425 has been
identified with an unusual supernova in ESO~184-G82 at a redshift
of $z=0.0085$ implying an energy output of $10^{52}$~erg
\cite{Galama98}.  These high energy outputs, combined with the
short duration and rapid variability on time-scales of
milliseconds, require highly relativistic motion to allow the MeV
photons to escape without severe photon-photon pair production
losses.  The energy sources of GRB may be neutron star mergers
with neutron stars or with black holes, collapsars associated
with supernova explosions of very massive stars, hyper-accreting
black holes, hypernovae, etc. (see \cite{Popham98,Iwamoto98} for
references to these models).

The~relativistic~fireball model of GRB (Meszaros and Rees
\cite{MesRees}) provides the framework for estimation of neutrino
fluxes from GRB.  A relativistic fireball sweeps up mass and
magnetic field, and electrons are energized by shock acceleration
and produce the MeV $\gamma$-rays by synchrotron radiation.  Protons
will also be accelerated, and may interact with the MeV
$\gamma$-rays producing neutrinos via pion photoproduction and
subsequent decay at energies above $\sim 10^{14}$~eV
\cite{WaxmanBahcall97,RachenMeszaros98}.  Acceleration of protons
may also take place to energies above $10^{19}$~eV, producing a
burst of neutrinos at these energies by the same process
\cite{Vietri98}.  These energetic protons may escape from the
host galaxy to become the highest energy cosmic rays
\cite{Waxman95,Vietri95}.  Additional neutrinos due to
interactions of the highest energy cosmic rays with the CMBR will
be produced as discussed in the previous section.

\begin{figure}[htb]
\vspace{7.1cm}
\includegraphics{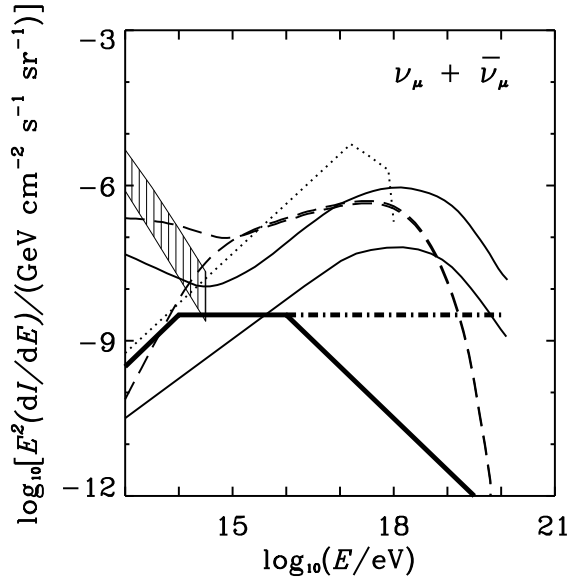}
\caption{Diffuse neutrinos from GRB: Thick solid line Waxman \&
Bahcall \protect\cite{WaxmanBahcall97}, with extension to higher
energies suggested by Vietri~\protect\cite{Vietri98} (thick chain
line).  Diffuse $p \gamma$ (lower) and $pp$ + $p \gamma$ (upper)
neutrinos from blazars: -------- Mannheim
\protect\cite{Mannheim95} ($p\gamma$) Model A (lower curve),
($p\gamma$+$pp$) Model B (upper curve); -- -- -- -- -- Protheroe
\protect\cite{Protheroe96} $\times 25$\% -- see text; $\cdots
\cdots$ $p \gamma$ Halzen \& Zas \protect\cite{HalzenZas97}.
\label{fig:bl}}
\end{figure}

For a sufficiently intense GRB, it may be possible to identify
neutrinos from individual GRB.  Integrating over all GRB in the
Universe, Waxman and
Bahcall~\cite{WaxmanBahcall97,WaxmanBahcall98} have predicted the
diffuse neutrino intensity, and this has been plotted in
Fig.~\ref{fig:bl} with a steepening at $10^{16}$~eV, and with a
continuation to higher energies as suggested by
Vietri~\cite{Vietri98}.

\section{ACTIVE GALACTIC NUCLEI}

The 2nd EGRET catalog of high-energy $\gamma$-ray sources
\cite{Thompson95} contains over 40 high confidence
identifications of AGN, and all appear to be blazars (radio-loud
AGN having emission from a relativistic jet closely aligned to
our line of sight).  Since the publication of the 2nd 
catalog, the number of blazars detected by EGRET has increased to
nearly 70 (see refs.~\cite{Montigny95,Mukherjee97} for reviews).
TeV emission has been observed from three blazars, the BL Lac
Objects Mrk~421, Mrk~501 and 1ES 2344+514 \cite{Catanese98}.
Clearly, the $\gamma$-ray emission is associated with AGN jets.
Blazars appear also to be able to explain about 25\% of the
diffuse $\gamma$-ray emission \cite{Chiang98}, and models where
$\gamma$-ray emission does not originate in the jet are unlikely
to contribute significantly to the diffuse $\gamma$-ray (and
neutrino) intensity (see Protheroe and Szabo
\cite{ProtheroeSzabo94} and references therein for predictions
for non-blazar AGN).  Several of the EGRET AGN show $\gamma$-ray
variability with time scales of $\sim 1$ day \cite{Kniffen93} at
GeV energies, and variability on time scales of $\sim 1$ hour or
less \cite{Gaidos96,ProtheroeMrk501} has been observed at $\sim 1$ TeV
for some BL Lacs.  These variability timescales place
important constraints on the models, and not all models developed
so far are consistent with this.  I shall survey the neutrino
emission predicted in blazar models irrespective of this,
assuming they may be made to accommodate the latest variability
measurements.

Most theoretical work on $\gamma$-ray emission in AGN jets
involved electron acceleration and inverse Compton scattering,
and these models will predict no neutrinos.  In proton blazar
models, protons are accelerated instead of, or as well as,
electrons.  In this case interactions of protons with matter or
radiation would lead to neutrino production.  In some of the
proton blazar models energetic protons interact with radiation
via pion photoproduction (see e.g. \cite{ProtheroeERICE96CR} for
references and a discussion of $p \gamma$ interactions).  This
radiation may be reprocessed or direct accretion disk radiation
\cite{Protheroe96}, or may be produced locally, for example, by
synchrotron radiation by electrons accelerated along with the
protons \cite{MannheimBiermann92,Mannheim95}.  Pair synchrotron
cascades initiated by photons and electrons resulting from pion
decay give rise to the emerging spectra, and this also leads to
quite acceptable fits to the observed spectra.  These models can
produce neutrinos and also higher energy radiation than electron
models because protons have a much lower synchrotron energy loss
rate than electrons for a given magnetic environment.  In both
classes of model, shock acceleration has been suggested as the
likely acceleration mechanism (see \cite{ProtheroeERICE96CR} for
references).

By appropriately integrating over redshift and luminosity in an
expanding universe, using a luminosity function (number density
of objects per unit of luminosity) appropriate to blazars, and
using the proton blazar models to model the $\gamma$ ray and
neutrino spectra one can estimate the diffuse neutrino background
expected from blazars.  In Fig.~\ref{fig:bl} I have added
intensities of $(\nu_\mu + \bar{\nu}_\mu)$ predicted in proton
blazar models by Mannheim \cite{Mannheim95}, Protheroe
\cite{Protheroe96} ($\times 0.25$ as only $\sim$25\% of
$\gamma$-ray background is due to AGN~\cite{Chiang98} -- original
calculation assumed 100\%) and Halzen and Zas \cite{HalzenZas97}.
For some of these models expected muon rates have been calculated
\cite{GaryHill96,Gandhi96,Gandhi98}.

\begin{figure}[thb]
\vspace{5.7cm}
\includegraphics{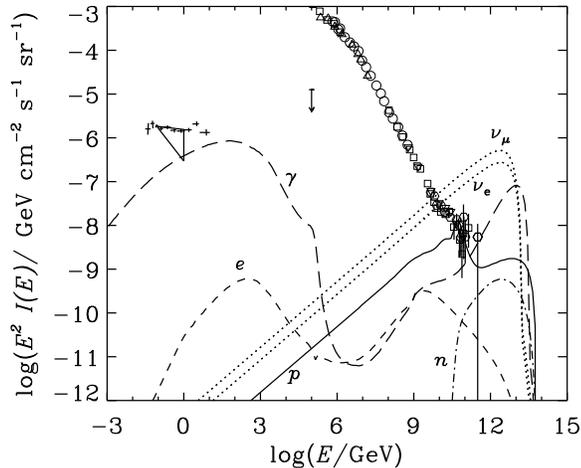}
\caption{The result of \protect\cite{ProtheroeStanev96} for
$M_Xc^2 = 10^{14.1}$ GeV, a magnetic field of $10^{-9}$ gauss,
and $p=2$, normalized the spectrum of ``observable particles''
(nucleons, photons, electrons) to the $3 \times 10^{11}$ GeV
data~\protect\cite{Bir95}).  $H_0=75$ km
s$^{-1}$ Mpc$^{-1}$ and $q_0=0.5$ are assumed.
\label{fig:td}}
\end{figure}

\section{TOPOLOGICAL DEFECTS}

Finally, I discuss perhaps the most uncertain of the components
of the diffuse high energy neutrino background, that due to
topological defects (TD).  In a series of papers
\cite{Hill83,AharonianBhatSchramm92,BhatHillSchramm92,GillKibble94},
TD have been suggested as an alternative explanation of the
highest energy cosmic rays.  In this scenario, the observed
cosmic rays are a result of top-down cascading, from somewhat
below (depending on theory) the GUT scale energy of $\sim
10^{16}$ GeV \cite{Amaldi91}, down to $10^{11}$ GeV and lower
energies.  These models put out much of the energy in
a very flat spectrum of neutrinos, photons and electrons
extending up to the mass of the ``X--particles'' emitted.

\begin{figure*}[hbt]
\vspace{7.1cm}
\includegraphics{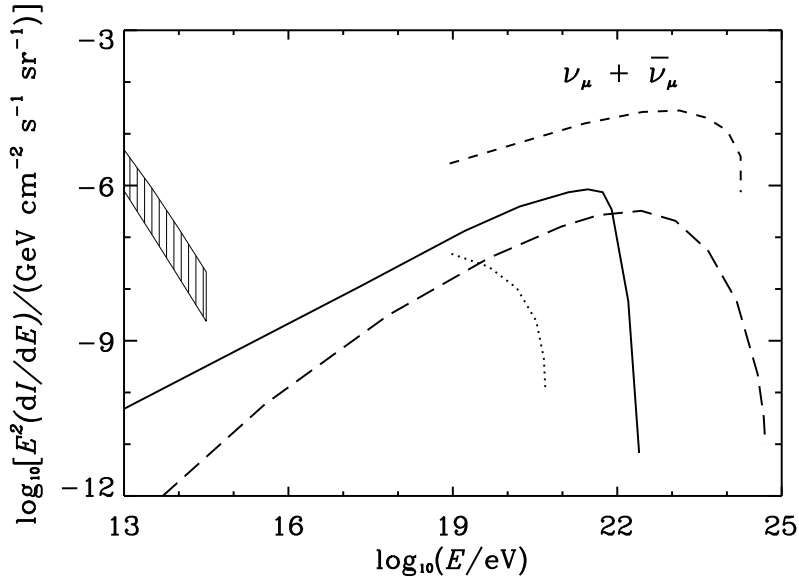}
\caption{Neutrinos from topological defects: -------- TD model
just ruled out according to Protheroe \& Stanev
\protect\cite{ProtheroeStanev96} ($10^{14.1}$ GeV,
$10^{-9}$~gauss); ---~---~---~---~ TD model just allowed according
to Sigl et al. \protect\cite{SiglLeeSchrammCoppi96} ($2 \times
10^{16}$ GeV, $10^{-12}$~gauss); $\cdots \cdots$ Birkel \& Sarkar
\protect\cite{BirkelSarkar98} ($10^{12}$ GeV, $0$~gauss);
-~-~-~-~- Yoshida et al.\protect\cite{Yoshida97} ($10^{16}$
GeV, $0$~gauss).
\label{fig:ex}}
\end{figure*}

Protheroe and Stanev \cite{ProtheroeStanev96} argue that these
models appear to be ruled out by the GeV $\gamma$-ray intensity
produced in cascades initiated by X-particle decay for GUT scale
X-particle masses.  The $\gamma$-rays result primarily from
synchrotron radiation of cascade electrons in the extragalactic
magnetic field.  Fig.~\ref{fig:td}, taken from
ref.~\cite{ProtheroeStanev96}, shows the neutrino emission for a
set of TD model parameters just ruled out according to Protheroe
and Stanev \cite{ProtheroeStanev96} for a magnetic field of
$10^{-9}$~G and X-particle mass of $1.3 \times 10^{14}$~GeV.
Clearly for such magnetic fields and higher X-particle masses
(e.g. GUT scale), TD cannot explain the highest energy cosmic
rays.  Indeed there is evidence to suggest that magnetic fields
between galaxies in clusters could be as high as $10^{-6}$~G
\cite{Kronberg94}.  However, for lower magnetic fields and/or
lower X-particle masses the TD models might explain the highest
energy cosmic rays without exceeding the GeV $\gamma$-ray limit.
For example, Sigl et al.~\cite{SiglLeeSchrammCoppi96} show that a
TD origin is not ruled out if the extragalactic field is as low
as $10^{-12}$~G, and Birkel \& Sarkar~\cite{BirkelSarkar98} adopt
an X-particle mass of $10^{12}$~GeV.  Yoshida et
al.~\cite{Yoshida97} investigate various TD scenarios with GUT
scale masses, and their predicted neutrino fluxes are generally
higher than those of Sigl et al.~\cite{SiglLeeSchrammCoppi96},
but such high neutrino intensities are likely to be excluded
because the $\gamma$-rays, due to cascading even in a $10^{-12}$~G
field, would probably exceed the GeV flux.  The intensities are
compared in Fig.~\ref{fig:ex}.  A novel feature of the work of
Yoshida et al.~\cite{Yoshida97} is the inclusion of interactions
of high energy neutrinos with the 1.9~K cosmic neutrino
background, and this can be important at the very highest
energies.

I emphasize that the predictions summarized in Fig.~\ref{fig:ex}
are {\em not} absolute predictions, but the intensity of
$\gamma$-rays and nucleons in the resulting cascade is normalized
in some way to the highest energy cosmic ray data.  It is my
opinion that GUT scale TD models are neither necessary nor able
to explain the highest energy cosmic rays without violating the
GeV $\gamma$-ray flux observed.  The predicted neutrino
intensities are therefore {\em extremely} uncertain.
Nevertheless, it is important to search for such emission
because, if it is found, it would overturn our current thinking
on the origin of the highest energy cosmic rays and, perhaps more
importantly, our understanding of the Universe itself.

\section{DISCUSSION}

Very recently, Waxman and Bahcall \cite{WaxmanBahcall98} have
used some arguments based on the observed cosmic ray spectrum to
obtain an upper bound to high energy neutrinos from astrophysical
sources.  Their argument hinges on sources of astrophysical
neutrinos being sources of the highest energy cosmic rays 
which happen also to produce neutrinos by $p\gamma$ interactions.
Hence, except for sources with a very high optical depth for
protons, the maximum neutrino intensity will be about 10\% of the
extragalactic ($\sim E^{-2}$) component of the highest energy
cosmic rays.  Examining AGN models, they find that
predictions for proton blazar models exceed their bound.

\begin{figure}[hbt]
\vspace{5.1cm}
\includegraphics{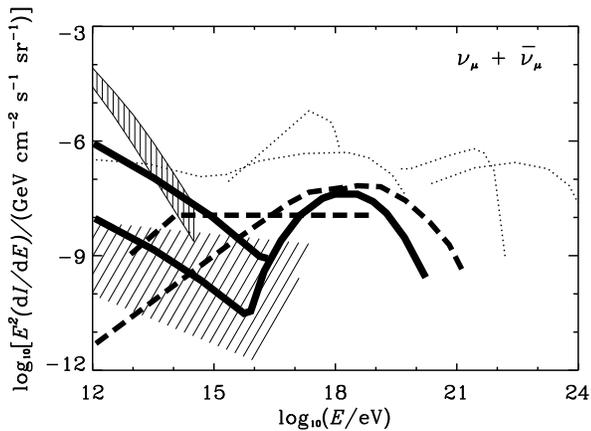}
\caption{Grand Unified Neutrino Spectrum -- a personal opinion
about the predicted neutrino intensities: thick solid lines --
certain; long dashed lines -- almost certain; short dashed lines
-- speculative; dotted lines -- highly speculative.
\label{fig:final}}
\end{figure}

In the case of AGN, they also suggest that the optical depth to
$p\gamma$ at $\sim 10^{19}$~eV must be much less than 1 (to enable
TeV $\gamma$-rays to escape without significant $\gamma \gamma$
pair production losses), with the consequence that the ultra high
energy cosmic ray production far exceeds the ultra high energy
neutrino production, pushing the neutrino upper bound even lower.  

TeV $\gamma$-rays, however, have so far only been seen from 3
blazars, and it is by no means certain that TeV $\gamma$-rays are
emitted by all blazars and so high $p\gamma$ optical depths are
not necessarily ruled out (note, however, that the infrared
background limits how far away one can observe objects at TeV
energies~\cite{Stecker98}).  Also, in at least one of the proton
blazar models \cite{Protheroe96} the optical depth of protons to
$p \gamma$ at $10^{19}$~eV is high because the proton directions
are isotropic in the jet frame whereas the radiation field is
highly anisotropic, coming from near the base of the jet, and the
photons cascade down to TeV energies where the $\gamma \gamma$
optical depth along the jet direction is low because of the
radiation being anisotropic.  Admitedly, neutrons are produced in a
fraction of $p \gamma$ interactions, and the neutrons escape as
cosmic rays, and so the effective optical depth for nucleons can
not exceed $\sim 1$ by much, and so it is probable that this proton
blazar model is ruled out.

The main argument relating the neutrino upper bound to the
observed ultra high energy cosmic ray flux relies on the cosmic
rays of energy $10^{19}$~eV being able to reach Earth from AGN
during the Hubble time.  There is evidence to suggest that
magnetic fields between galaxies in cores of clusters (the most
likely place to find an an AGN) could be as high as $10^{-6}$~G
\cite{Kronberg94}.  With such high magnetic fields it is not
obvious that $10^{19}$~eV protons will reach us from most AGN
contributing high energy neutrinos.

Thus, I believe that the ``upper bound'' is model dependent, and
that its calculation is complicated by cosmic ray propagation
effects.  While I would certainly classify the higher AGN fluxes as
speculative, or highly speculative, I believe the lower ones are
not ruled out by the argument of Waxman and Bahcall.
Nevertheless, the work of Waxman and Bahcall is very important in
reminding us that for any model used to predict high
energy neutrino fluxes we must check that it does not overproduce
cosmic rays.

Plotting a representative sample of the diffuse flux predictions
from Figs.~\ref{fig:cr3}, \ref{fig:cr2}, \ref{fig:bl} and
\ref{fig:ex} in the same figure one has a ``grand unified
neutrino spectrum'' (with apologies to Ressell and Turner
\cite{RessellTurner91}).  This is shown in Fig.~\ref{fig:final}
where I have labelled the various curves as ``speculative'',
``highly speculative'', ``certain'' or ``almost certain''.  These
labels reflect my own personal opinion or prejudice and should not
be taken too seriously -- other opinions are equally valid.

\section{PROSPECTS FOR OBSERVATION}

With the construction in the relatively short term of 0.1~km$^2$
neutrino telescopes, and in the longer term of 1~km$^2$
detectors, it is useful to estimate the signals expected due to
various possible neutrino intensities.  At high energies,
electron neutrinos may also be detected through the resulting
cascade, and this is particularly important when looking for
horizontal air showers, for example with the proposed AUGER
detectors~\cite{Yoshida97}.  Several estimates of event rates
have been made for various energy thresholds, or for horizontal
air showers due to neutrino interactions (including $\nu_e$ and
$\bar{\nu}_e$, see e.g. Gandhi et al.~\cite{Gandhi98}).

\begin{figure}[t]
\vspace{5.7cm}
\includegraphics{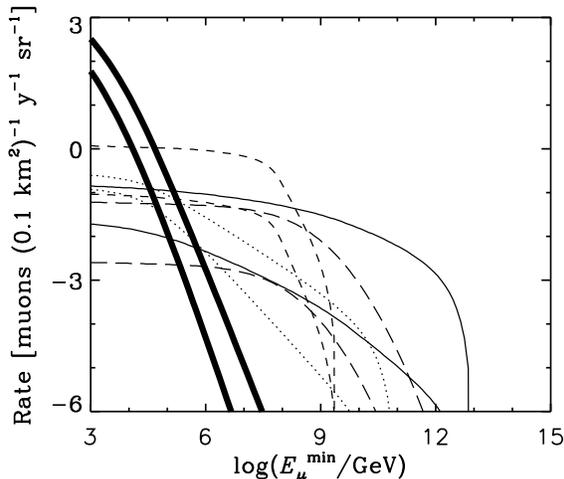}
\caption{Atmospheric neutrino induced muon signal (thick solid
lines) and putative astrophysical neutrino induced muon signals
expected for the following ($\nu_\mu + \bar{\nu}_\mu$)
intensities: GRB intensity~\protect\cite{WaxmanBahcall97}
extended to $10^{20}$~eV (dotted curves); $p \gamma$ proton
blazar~\protect\cite{Mannheim95} (short dashed curves); TD model just
ruled out according to ref.~\protect\cite{ProtheroeStanev96}
(solid curves); interactions of the highest energy cosmic rays
with the CMBR \protect\cite{ProtheroeJohnson95} for $E_{\rm
max}=3 \times 10^{21}$~eV (long dashed curves).  Upper curves
show horizontal signals, lower curves show vertical (upward)
signals.
\label{fig:signal}}
\end{figure}

To illustrate how the ($\nu_\mu + \bar{\nu}_\mu$)
signals expected from different astrophysical neutrino spectra
would be detected by telescopes with different energy thresholds,
I have made approximate estimates of the event rates {\em as a
function of minimum muon energy} using the $P_{\nu \to
\mu}(E_\nu, E_\mu^{\rm min})$ function given in Fig.~2 of
ref.~\cite{GaisserHalzenStanev95} for $E_\mu^{\rm min} = 1$~GeV,
modified for other $E_\mu^{\rm min}$ values in a way consistent
with that given for $E_\mu^{\rm min} = 1$~TeV.  The effects of
shadowing for vertically upward-going neutrinos have been
included using the shadow factor $S(E_\nu)$ given in Fig.~20 of
ref.~\cite{Gandhi96}.  I have estimated the expected neutrino
induced muon signal for four representative neutrino intensities.
The vertically upward-going and horizontal muon signals are shown
separately for each case in Fig.~\ref{fig:signal} together with
the atmospheric neutrino induced muon signals for the two
directions.  As can be seen, the highest signals would be due to
the proton blazar models, with several events per year expected
in a 0.1~km$^2$ detector.  However, one should be cautious as
these intensities are somewhat speculative (as discussed
earlier).  Detection of muon signals in one year from the other
intensities estimated would be marginal for a 0.1~km$^2$
detector, but achievable with a 1~km$^2$ detector.  Detection of
transient neutrino signals, correlated with observations of the
same source in photons (e.g. GRB, AGN) should therefore be the
goal of high energy neutrino astronomy in the short term.

One should consider the consequences for astrophysical neutrinos
of the discovery of the oscillation of atmospheric $\nu_\mu$,
probably into $\nu_\tau$, by Super-Kamiokande \cite{SuperK98}
with an oscillation length of $\lambda_{\rm osc} \sim
10^3$($E$/GeV)~km.  On an astrophysical scale, the oscillation
length $\lambda_{\rm osc} \sim 3 \times 10^{-11}$($E$/TeV)~kpc is
very small, and integrating contributions to the neutrino
intensity over astrophysical dimensions one would naively expect
the ($\nu_\mu + \bar{\nu}_\mu$) flux to be 50\% lower (assuming
$\sin^22\theta=1$), and to be accompanied by a similar ($\nu_\tau
+ \bar{\nu}_\tau$) flux.  The unique signature for detection of
tau neutrinos has been discussed in ref.~\cite{LearnedPakvasa95}.

Neutrino astronomy is developing during an era in which exciting
discoveries are being made in other areas of high energy
astrophysics.  These include detection of rapidly varying TeV
$\gamma$-ray signals from AGN, discovery that GRB are
extragalactic and probably the most energetic phenomena
ocurring in the Universe today, and detection at Earth of
cosmic rays with energies well above $10^{20}$~eV opening the
question of whether their origin is through particle acceleration
at radio galaxies or GRB, or from topological defects left over
from the big bang.  Hadronic processes may have a role in all
these phenomena, and searching for high energy neutrinos may lead
to greater understanding of the highest energy phenomena in the
Universe.  Clearly, for this to happen in parallel with the other
observations rapid development of neutrino telescopes with
sensitive areas of $\sim 1$~km$^2$ or larger, operating over a
wide range neutrino energies, is essential.

\section*{ACKNOWLEDGMENTS}

I thank A. M\"{u}cke and Q. Luo for reading the manuscript, and
J. Bahcall and E. Waxman for helpful comments.  This research is
supported by a grant from the Australian Research Council.

\end{document}